# Comments to the problem of experimental determination of the neutron-electron scattering length and its theoretical interpretation.

A.B. Popov[a], T.Yu. Tretyakova

Joint Institute for Nuclear Research, 141980 Dubna, Russia

**Abstract.** We discuss the experimental data on the n,e-scattering length $b_{ne}$ and the values of mean square charge radius of the neutron $<r_e^2>_n$ obtained from them. It is shown that the accumulated during the last 50 years most significant experimental estimates of the $b_{ne}$ are not contradictory and lead to the average value
$$<r_e^2> = -0.1178 \pm 0.0037 \; fm^2.$$
Assuming that all the authors have underestimated the errors of their measurements by a factor of 1.7, the combined fit of all available experimental data would lead to $\chi^2 \sim 1$ per degree of freedom. Different modern theoretical predictions of $<r_e^2>_n$ are considered. They are found to be in a good agreement with the obtained experimental value $<r_e^2>_n$. However the existing theoretical description of the structure of neutron does not provide a value of $<r_e^2>_n$ with a sufficient accuracy.

## 1 Introduction

Since first work of E. Fermi [1] for more than 50 years the question about n,e-interaction attracts attention of both experimentalists and theorists. During these years many experiments with different methods for determination of n,e-scattering length $b_{ne}$ were carried out and theoretical ideas about connection of $b_{ne}$ with mean square charge radius of a neutron $<r_e^2>_n$ were developed. The view of nucleon internal structure has changed significantly and now it is based on Standard Model principles. Unfortunately, the existing experimental estimates of $b_{ne}$ are widely different, which allows to divide these values in two groups which differ by more than three standard deviations [2, 3, 4, 5]. Moreover, in the beginning of 50$^{th}$ Foldy showed that n,e-scattering length $b_{ne}$ could be divided in two parts, the second one depended on anomalous magnetic moment of neutron and named as "Foldy length"
$$b_F = \frac{\mu e^2}{2M_n c^2} = -1.468 \cdot 10^{-3} \; fm \quad [6].$$
Taking into account this Foldy length one can obtain so-called "intrinsic" charge radius of neutron $<r_1^2>$, which will have different sign for the two groups of experimental results. The physical meaning of $<r_1^2>$, at least its sign, was the subject of a long-term discussion. The positive value was admitted by some authors as unphysical {3}, that put under doubt results of the most numerous group of experiments, which gave values of n,e-scattering length
$$b_{ne} = -(1{,}32 \pm 0.03) \cdot 10^{-3} \; fm \, .$$
In work [7] the Foldy's description of n,e-scattering length was revised on the basis of the Dirac's equation and it was emphasized that $b_{ne}$ is related to the complete coefficient in front of $div\overline{E}$ - term in the equation, so that the charge distribution is connected with the total value of $b_{ne}$ only. Consequently the mean square charge radius of a neutron is related only to this value, i.e.
$$<r_e^2>_n = \frac{3\hbar^2}{M_n e^2} b_{ne}. \qquad (1)$$
Since papers [8] and [9] it is accepted to estimate the value of $<r_e^2>_n$ using this equation. The coefficient before $b_{ne}$ can be expressed through the other constants: $\frac{3m_e}{M_n} a_0$ (where $a_0 = \frac{\hbar^2}{m_e e^2}$ is the Bohr radius) or $\frac{3\hbar c}{\alpha M_n c^2}$ ($\alpha = \frac{e^2}{\hbar c}$).

In the compilation of the Particle Data Group [10] the ten experimental results are shown and the recommended value is obtained by averaging only five of them:
$$<r_e^2>_n = -0.1161 \pm 0.0022 \; fm^2. \qquad (2)$$
In this compilation the authors use the result of [11], obtained in 1986 from the neutron total cross section of Bi, which was subjected to serious criticism in [12]. On the other hand the largest absolute value of $b_{ne} = -1.60 \pm 0.05 \; mfm$ from neutron diffraction on a single crystal of $^{186}W$ [13] was ignored.

---
[a] e-mail: popov_ab@nf.jinr.ru

## 2 Analysis of existing measurements

In present work we consider a more complete set of $b_{ne}$ experimental data and corresponding values of $<r_e^2>_n$, adding estimates of $b_{ne}$ from works [13] (for a single crystal of $^{186}W$), [14], [15], [16] (corrected for Schwinger's scattering in [2]) and recent result [17], obtained from the structural factors for liquid $Kr$ by method proposed in [5].

All collected data are summarized in Table 1 and fig. 1. The calculations of the average $<r_n^2>$ value are performed using the MINUIT program [23] for several approaches. The corresponding results are shown in Table 2. We use different sets of experimental results: first 14 experiments from Table 1, the same set but without the value for $^{186}W$, the set with two additional points from new estimations of $b_{ne}$ from neutron scattering total cross section on $^{208}Pb$, obtained from the analysis of Garching group data [8] and the result of FLNP experiment [22]. In the last set we also use the original result [11] instead of re-estimated value of $<r_n^2>$ for $Bi$ from [20].

Thus, we can conclude from Table 2 that the existing estimates of $b_{ne}$ give an average value of $<r_n^2>$:

$$<r_e^2>_n = -0.1178 \pm 0.0037 \; fm^2 . \qquad (3)$$

In the fig. 1 we show the fitting results of average $<r_e^2>_n$ value using the original experimental data. It can be seen that all experimental points differ from the average value by not more than three standard deviations except the result for $^{186}W$ [13]. Rejection of this point does not change the result significantly and leads to the average value:

$$<r_e^2>_n = -0.1153 \pm 0.0024 \; fm^2 . \qquad (4)$$

Both results correspond to 95% confidential interval and are in a good agreement within errors with the value recommended by Particle Data Group.

It is possible to draw a conclusion, that all available experimental estimates of $b_{ne}$ ($<r_e^2>_n$) are not in contradiction with each other under assumption that the authors have underestimated errors of their experiments by less than a factor of two. This assumption is quite realistic taking into account the presence of strong corrections of a different origin in each experiment, which accuracy is really limited and requires a serious reassessment.

## 3 Theoretical aspects

Let us consider the modern theoretical view on neutron charge structure. First of all it should be noted that the definition of mean square charge radius $<r_e^2>_n$ works in coordinate space in non-relativistic assumption. But in most of modern theoretical approaches hadron charge structure is considered in impulse space and its distinctive characteristic is the dependence of nucleon electromagnetic form factors on momentum transfer $q$. At small $q$ form factors reflect such nucleon features as its charge and magnetic moment, radii of charge and magnetic moment distributions, whereas at large ones they constitute the information on quark structure of nucleon corresponding to quantum chromodynamics.

Electromagnetic structure of nucleon is determined by matrix element of current operator $j$, which can be expressed via two form factors:

$$\langle \mathbf{p}'|j_\mu|\mathbf{p}\rangle = \bar{u}(\mathbf{p}')[F_1(q^2)\gamma_\mu + F_2(q^2)i\sigma_{\mu\nu}q^\nu/2M]u(\mathbf{p}),$$

where $M$ is the nucleon mass, $q^2$ is the square of momentum transfer. $F_1(q^2)$ is Dirac form factor carried information on charge and normal magnetic moment of nucleon, $F_2(q^2)$ is the Pauli form factor corresponded to particle anomalous magnetic moment. Form factors $F_1(q^2)$ and $F_2(q^2)$ for proton and neutron are normalized at $q^2=0$ on their charge and magnetic moment values:

$$F_1^p(0) = 1, \; F_1^n(0) = 0, \; F_2^p(0) = 1{,}79, \; F_2^n(0) = -1{,}91 .$$

For description of the nucleon total magnetic moment and charge radius in [24] charge and magnetic form factors were introduced

$$G_E(q^2) = F_1(q^2) - \frac{q^2}{4M^2}F_2(q^2) \qquad \text{and} \qquad (5)$$

$$G_M(q^2) = F_1(q^2) + F_2(q^2). \qquad (6)$$

Such expressions are optimal from the experimental analysis point of view because they do not interfere in well known Roesenbluth formulae for differential cross section of electron scattering on space target with spin ½:

$$\frac{d\sigma}{d\Omega} = \left(\frac{d\sigma}{d\Omega}\right)_{Mott} \left\{ \frac{G_E^2 - \frac{q^2}{4M^2}G_M^2}{1 - \frac{q^2}{4M^2}} - \frac{q^2}{2M^2}G_M^2 tg^2\left(\frac{\Theta}{2}\right) \right\}$$

Sacks showed [25] that in Breit system just form factors of such form corresponded to Fourier transforms of the charge and magnetization distributions, that is why they were named as electric and magnetic ones. Formally all other form factors (Dirac, Pauli, their isovector or isoscalar combinations) also can be presented as Fourier transforms of some space distributions too, but most likely it would be formal presentations without any physical sense. Respectively for every form factor at $q^2 \to 0$ the mean square radius of associated space distribution can be defined:

$$<r_i^2> = -6 \frac{dF_i(q^2)}{dq^2}\bigg|_{q^2=0} .$$

In the case of charge form factor $G_E$ this quantity corresponds to nucleon mean square charge radius obtained from experiment.

As soon as $G_E$ is expressed through combination of Dirac and Pauli form factors, the neutron charge radius $<r_e^2>_n$ can be presented as a sum:

$$<r_e^2>_n = r_1^2 + r_{Foldy}^2, \qquad (7)$$

where $r_1^2 = -6 \dfrac{dF_1^n(q^2)}{dq^2}\bigg|_{q^2=0}$ and $r_{Foldy}^2 = \dfrac{3\mu_n}{M^2}$

($\mu_n = F_2^n(0)$ - the neutron anomalous magnetic moment). The second part, Foldy term [6], appears due to the generation of the electric field by the anomalous neutron magnetic moment because of its "zitterbewegung". With the neutron magnetic moment $\mu_n = -1.91$ the value of this term is $r_{Foldy}^2 = -0.126 \, fm^2$ and it is very close to experimental estimations of $<r_e^2>_n$.

This fact attracts the attention of theoreticians and during last years several works have evolved which consider the radius $r_1^2$, connected with Dirac form factor [26, 27, 28, 29, 30] and attempt to find the physical meaning of this quantity. It was shown that in non-relativistic approximation, with the SU(6) symmetric wave function, the neutron charge form factor is identically zero $G_E^n(q^2) \equiv 0$. Attempts to incorporate the relativistic effects result in $r_1^2 = -r_{Foldy}^2$, so in this approximation the value of neutron charge radius is still zero $<r_e^2>_n = 0$ [27, 28].

In [26] the higher order in expansion on $1/m^n$ (where $m$ is the quark mass) was used and the dependence of nucleon characteristics on quark anomalous magnetic moments was considered. The fit of the nucleon static properties results in the reasonable values for nucleon magnetic moments $\mu_n = -1,92$ and $\mu_p = 3,09$, and neutron charge radius $<r_e^2>_n = -0,125 \, fm^2$
($r_1^2 = 0,002 \, fm^2$, $r_{Foldy}^2 = -0,127 \, fm^2$)

The other authors [29] believe that the agreement between $<r_e^2>_n^{exp}$ and $r_{Foldy}^2$ is accidental and underline that the rest frame charge distribution of the neutron should be associated with the form factor $G_E^n$ and not with $F_1^n$. In this work it was shown that in non-relativistic regime the cancellation between $r_1^2$ and $r_{Foldy}^2$ happens indeed for large nucleon sizes and it is independent of the detailed form of quark spin coupling scheme and wave functions, while at the physical nucleon scale the value of $r_1^2$ is strongly dependent on the choice of quark spin coupling scheme.

In [28] the connection of neutron charge radius with Dirac equation is discussed more closely. Authors consider the Dirac equation for a finite-size neutron in an external electric field and incorporate Dirac-Pauli form factors in it explicitly. After a non-relativistic reduction, the Darwin-Foldy term is cancelled by a contribution from the Dirac form factor, so that the only coefficient of the external field charge density is $\dfrac{e}{6}<r_e^2>_n$, i. e. the mean square radius associated with the electric Sachs form factor $G_E^n$. This result is similar to a result of [27], however it is independent from any definite neutron quark substructure. The neutron just has to have a form factor. Furthermore, the analysis is in keeping with the philosophy that the basic equations for the neutron should be expressed in terms of a Dirac Hamiltonian, while the physical picture emerges from a non-relativistic reduction, which only contains $<r_e^2>_n$. In [28] it is noted that $G_E^n$ contains the buried term, which depends on neutron anomalous magnetic moment and contributes the most to the neutron charge radius. The analysis and conclusion are in good agreement with the analysis of [31] for low-energy Compton scattering from nucleon.

The results of these calculations and nucleon features obtained in other models are given in table 3.

It should be noted that we are interested in region of low momentum transfer, which is beyond the limits of perturbative quantum chromodynamics, so for nucleon static characteristics description some models should be used. The one exception is the lattice calculations share basic principles of QCD. During the past few years this field has progressed substantially and now the results for nucleon properties agree satisfactorily with the experimental data [36, 37]. However these calculations have not given an understanding the physical picture, the different model approaches as quark model, for example, retain their importance. A reliable calculation of the nucleon static characteristics should incorporate many contributions, such as relativistic effects in the nucleon wave function, its nontrivial spin structure, exchange currents within nucleon, pionic fluctuations of constituent quarks [30], anomalous quark magnetic moments, etc [26]. Calculations of these corrections are a difficult enough problem so it is difficult to treat the accuracy of one model or another and to wait for precise description of experimental data. At present state the experimental results are more accurate than theoretical calculations. They can be used as a criteria in deciding between different theoretical models [30].

Thus it can be stated that in recent publications (except [3]) the authors do not mention the separation of Foldy term from experimental value of $b_{ne}$. Obtained from $b_{ne}$ experimental estimate of $<r_e^2>_n$ is considered as a total mean square charge radius of neutron. The old statement [3] about division of experimental data in two groups and that part of these data are in contradiction to modern physical theories has no meaning evidently (see table 1 and figure). Considering theoretical approaches mentioned above one can notice that $<r_e^2>_n$ can be divided into two components, related to Dirac and Pauli form factors, but in different models the relation between these two components is different and so-called "intrinsic" charge radius of neutron $<r_1^2>$ connected with the Dirac form factor may have either positive or negative sign. Moreover in some papers it was shown that consideration of Dirac equation for neutron at non-relativistic limit leads to vanishing the Foldy term, while the term, expressed through $<r_e^2>_n$, is still present. It is responsible for interaction of neutron with external electric field. This means that in Born approximation the n,e-scattering length $b_{ne}$ depends only on the total charge radius of neutron $<r_e^2>_n$, as stated previously in [7].

## 4 Conclusion

The problem of $b_{ne}$ and neutron charge radius estimation appears to be less pressing. It is unlikely that existing set of experimental data should be considered as self-contradictory. The average value from all experiments

$$<r_e^2>_n = -0.1178 \pm 0.0037 \ fm^2$$

within limits of its accuracy of 4% corresponds to confidence interval of 95% even including the result for $^{186}W$, which differ from the average value by more than 5 standard deviations. If this measurement is excluded, the average value becomes

$$<r_e^2>_n = -0.1153 \pm 0.0024 \ fm^2 \ ,$$

with accuracy $<r_e^2>_n$ - 2.5% for confidence interval of 95%.

Of course, extraction of $b_{ne}^{exp}$ from different experiments required important corrections to be made. That is why new experimental proposals using new approaches are very interesting, however it is naive to expect that they could lead to abrupt changes in the problem considered. It is very important to perform an experiments with accuracy better than 2.5%. Proposals of precise measurements using interferometers are worth to mention [4, 38, 39]. The experiments to measure the structure factors of noble gases [5, 17] may be useful also. However more precise experimental determination of $<r_e^2>_n$ cannot improve its physical interpretation due to ambiguity of theoretical descriptions of nucleon structure nowadays. But this precise experimental value of $<r_e^2>_n$ would be very helpful in future development of nucleon characteristics.

**Table 1.** n,e-scattering length $b_{ne}$ and mean square charge radius of neutron $<r_n^2>$.

Bold font in column "Year" marks the value used in later averaging.

**Table 2.** Average values of $<r_e^2>_n$.

**Table 3.** Theoretical estimations of nucleon magnetic moments and mean square charge radii.

Fig. 1. Neutron mean square charge radius $<r_e^2>_n$ from different experiments and the average value. The shaded area shows the 95 % confidential interval.

Table 1

| Experiment | Year | $b_{ne} \cdot 10^{-3}$, fm | $<r_n^2>$, fm$^2$ |
|---|---|---|---|
| W. Havens, liquid Bi $\sigma_t$ [14] | **1951** | $-1.89 \pm 0.36$ | $-0.163 \pm 0.31$ |
| Huhges, mirror Bi/O$_2$ [15] | **1953** | $-1.39 \pm 0.13$ | $-0.120 \pm 0.011$ |
| Melkonian, Bi cryst spectr $\sigma_t$ [16] | 1959 | $-1.56 \pm 0.05$ | $-0.135 \pm 0.004$ |
| Re-estimation, Koester | 1976 | $-1.49 \pm 0.05$ | $-0.127 \pm 0.004$ |
| Re-estimation, Kopecky [2] | **1997** | $-1.44 \pm 0.03 \pm 0.06$ | $-0.124 \pm 0.003 \pm 0.008$ |
| Krohn, angle distribution on gases | 1966 | $-1.34 \pm 0.03$ | $-0.116 \pm 0.003$[**] |
| Re-estimation, Krohn [18] | **1973** | $-1.33 \pm 0.03$ | $-0.115 \pm 0.003$[*] |
| Alexandrov, $^{186}W$ [13] | **1975** | $-1.60 \pm 0.05$ | $-0.138 \pm 0.004$ |
| Koester, filters - $b_{coh}$ Pb [19] | 1976 | $-1.364 \pm 0.025$ | $-0.118 \pm 0.002$[**] |
| Re-estimation, Nikolenko [20] | **1990** | $-1.32 \pm 0.03$ | $-0.114 \pm 0.003$ |
| Koester, filters - $b_{coh}$ Bi [19] | 1976 | $-1.393 \pm 0.025$ | $-0.120 \pm 0.002$[**] |
| Re-estimation, Nikolenko [20] | **1990** | $-1.33 \pm 0.03$ | $-0.115 \pm 0.003$ |
| Alexandrov, TOF $\sigma_t$ - $b_{coh}$ Bi [11] | 1986 | $-1.55 \pm 0.11$ | $-0.134 \pm 0.009$[*] |
| Re-estimation, Nikolenko [20] | **1990** | $-1.40 \pm 0.04$ | $-0.121 \pm 0.004$ |
| Koester, filters - $b_{coh}$ Pb, Bi [21] | **1986** | $-1.32 \pm 0.04$ | $-0.114 \pm 0.003$[**] |
| Kopecky, liquid $^{208}Pb$ TOF $\sigma_t$ [9] | **1995** | $-1.31 \pm 0.03 \pm 0.4$ | $-0.113 \pm 0.002 \pm 0.003$[**] |
| Koester, $^{208}Pb$, $Bi$ TOF [8] | **1995** | $-1.32 \pm 0.03$ | $-0.114 \pm 0.003$[*] |
| Kopecky, liquid $^{208}Pb$ TOF $\sigma_t$ [2] | **1997** | $-1.33 \pm 0.03 \pm 0.03$ | $-0.115 \pm 0.003 \pm 0.003$[*] |
| Kopecky, liquid Bi TOF $\sigma_t$ [2] | **1997** | $-1.44 \pm 0.03 \pm 0.04$ | $-0.124 \pm 0.003 \pm 0.005$[*] |
| Magli, diffraction on liquid Kr [17] | **2006** | $-1.40 \pm 0.10$ | $-0.121 \pm 0.009$ |
| Wasch.-LNP, $^{208}Pb$ | **2006** | $-1.56 \pm 0.18$ | $-0135 \pm 0.016$ |
| LNP, $^{208}Pb$ TOF $\sigma_t$-old [22] | **2006** | $-1.70 \pm 0.15$ | $-0.147 \pm 0.013$ |
| Estimation of Particle Data Group [10] | 2006 | | $-0.1161 \pm 0.0022$ |

[*] Data used for estimation of average $<r_n^2>$ by Particle Data Group

[**] Data included in Table of Particle Data Group but not used for estimation of $<r_n^2>$

**Table 2.** Average values of $<r_e^2>_n$.

| Number of points | $\chi^2$ | $<r_e^2>_n, fm^2$ | Confidence interval |
|---|---|---|---|
| 14 points | 39.4 | $-0.1172 \pm 0.0012\,(0.0021^*)$ | 67% |
| 14 points | 39.4 | $-0.1172 \pm 0.0023\,(0.0040^*)$ | 95% |
| Without $^{186}W$ | 9.87 | $-0.1153 \pm 0.0012$ | 67% |
| - "- | 9.87 | $-0.1153 \pm 0.0024$ | 95% |
| With two last points for $^{208}$Pb and Bi from [11], 17 points | 46.5 | $-0.1178 \pm 0.0022\,(0.0037^*)$ | 95% |
| 17 points, errors are increased by $\times \sqrt{\dfrac{\chi^2}{n-1}}$ | 16.1 | $-0.1178 \pm 0.0037$ | 95% |

$^*$**Error corrected by factor** $\times \sqrt{\dfrac{\chi^2}{n-1}}$

**Table 3.** Theoretical estimations of nucleon magnetic moments and mean square charge radii.

|  | Reference | $\mu_p$ | $\mu_n$ | $<r_e^2>_p$, fm$^2$ | $<r_e^2>_n$, fm$^2$ | $<r_1^2>$, fm$^2$ |
|---|---|---|---|---|---|---|
| **Experiment** |  |  | -1.91 |  | -0.118 |  |
| PDG | [10] | 2.79 | -1.91 | 0.757(14) | -0.1161(22) |  |
| **Model** |  |  |  |  |  |  |
| RQM | [30] |  |  |  |  | -.005÷0.009 |
|  | [32] | 2.88 | -1.58 | 0.62 | -0.185 | -0.084 |
|  | [26] | 3.09 | -1.92 |  | -0.125 | 0.002 |
| QM | [33] | 3.05 | -1.55 | 0.58 | -0.256 |  |
| PQCD | [34] |  |  | 0.689 | -0.119 |  |
| LFCBM | [35] | 2.95 | -1.79 |  | -0.110 |  |
| LC | [36] |  |  |  | -0.113 (17) |  |
|  | [37] | 2.72 (26) | -1.82 (34) | 0.685 (47) | -0.158 (29) |  |

Fig.1.

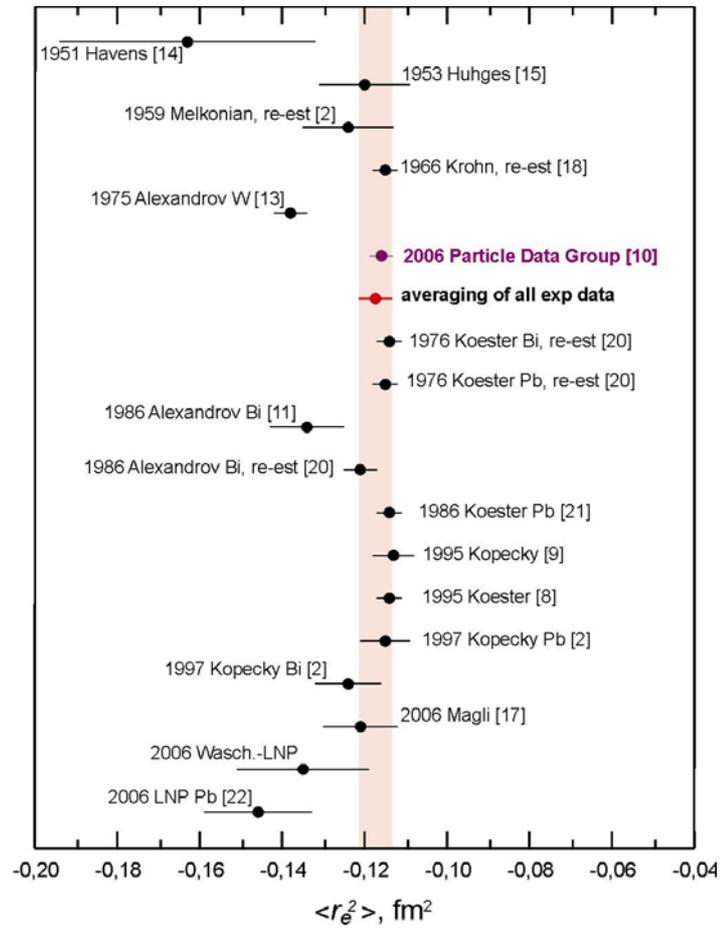